\documentclass[pra,twocolumn,notitlepage,floatfix,superscriptaddress]{revtex4-2}

\usepackage[T1]{fontenc}
\usepackage[colorlinks=true,urlcolor=blue,citecolor=blue,anchorcolor=red]{hyperref}
\usepackage{graphicx,url,color,physics,bbm,soul,mathtools,amssymb,amsmath,dsfont,lmodern,mathrsfs}

\begin{document}

\title{Filter-enhanced adiabatic quantum computing on a digital quantum processor}

\author{Erenay Karacan}
\email{ekaracan@ethz.ch}
\affiliation{Quantinuum, Leopoldstra{\ss}e 180, 80804 M\"{u}nchen, Germany}
\author{Conor Mc Keever}
\affiliation{Quantinuum, Partnership House, Carlisle Place, London SW1P 1BX, United Kingdom}
\author{Michael Foss-Feig}
\affiliation{Quantinuum, 303 S.\ Technology Ct., Broomfield, Colorado 80021, USA}
\author{David Hayes}
\affiliation{Quantinuum, 303 S.\ Technology Ct., Broomfield, Colorado 80021, USA}
\author{Michael Lubasch}
\email{michael.lubasch@quantinuum.com}
\affiliation{Quantinuum, Partnership House, Carlisle Place, London SW1P 1BX, United Kingdom}

\date{July 18, 2025}

\begin{abstract}
Eigenstate filters underpin near-optimal quantum algorithms for ground state preparation.
Their realization on current quantum computers, however, poses a challenge as the filters are typically represented by deep quantum circuits.
Additionally, since the filters are created probabilistically, their circuits need to be rerun many times when the associated success probability is small.
Here we describe a strategy to implement a ground-state filter on quantum hardware in the presence of noise by prepending the filter with digitized adiabatic quantum computing.
The adiabatically prepared input state increases the success probability of the filter and also reduces its circuit depth requirements.
At the same time, the filter enhances the accuracy of the adiabatically prepared ground state.
We compare the approach to the purely adiabatic protocol through numerical simulations and experiments on the Quantinuum H1-1 quantum computer.
We demonstrate a significant improvement in ground-state accuracies for paradigmatic quantum spin models.
\end{abstract}

\maketitle

\section{Introduction}

Ground state calculations on quantum computers are useful to numerous areas of research, e.g.\ quantum chemistry~\cite{SzOs96, AsEtAl05, SiVaRe19, CaEtAl19, McEtAl20, Ba20, LiEtAl22, LeEtAl23}, condensed matter physics and materials science~\cite{IbLu09, Ma10, Le13, Ba20}.
Over the past few years, considerable research efforts have focused on variational quantum algorithms~\cite{CeEtAl21a, BhEtAl22} for ground state computations~\cite{TiEtAl22}, due to the feasibility of running them on current quantum hardware.
Although substantial progress has been made using variational approaches, skepticism persists because of their heuristic nature and an ongoing dispute over their scalability~\cite{McEtAl18, CeEtAl21b, UvBi21, ZhGa21, PaEtAl21, OrKiWi21, WaEtAl21, HoEtAl22, LaEtAl22, CePlLu23}.
These aspects motivate a renewed interest in non-variational quantum algorithms that usually come with asymptotic performance guarantees but are generally too demanding to realize on present quantum devices.

One particularly versatile, non-variational procedure is adiabatic quantum computing (AQC)~\cite{KaNi98, FaEtAl00, AlLi18}, which starts from an easy-to-prepare ground state $\ket{\psi_{\text{init}}}$ of an initial Hamiltonian $H_{\text{init}}$ and transforms it into a final state $\ket{\psi_{\text{AQC}}}$ that approximates the desired ground state $\ket{\psi_{\text{targ}}}$ of the target Hamiltonian $H_{\text{targ}}$.
This transformation is governed by a time-dependent Schr\"{o}dinger equation that evolves the state from an initial time $t = 0$ to a total time $t = T$ according to the interpolating Hamiltonian
\begin{equation}\label{eq:adiabatic_ham}
 H(t) = \left[1-s(t)\right] H_{\text{init}} + s(t)H_{\text{targ}}.
\end{equation}
Here $s(t)$ parametrizes the so-called adiabatic path with $s(0) = 0$ and $s(T) = 1$.
The target ground state is successfully prepared if $T$ is sufficiently large and, often, $T$ needs to scale as $O\left(1/\Delta_{\text{min}}^{2}\right)$ where $\Delta_{\text{min}}$ is the minimum spectral gap (i.e.\ energy difference between the ground and first-excited state) of all Hamiltonians along the adiabatic path.
While AQC originally assumes evolution in continuous time, a digitized AQC form~\cite{FaEtAl00, VaMoVa01, BoKnSo09, BaEtAl16} has been developed for gate-based quantum computers with exciting applications~\cite{SuSoOr19, CoEtAl22, CoEtAl24}.
If highly accurate ground state preparation is required, however, the conditions for adiabaticity can lead to large circuit depths, which in turn impose challenging experimental requirements on quantum gate errors.

Quantum algorithms based on eigenstate filtering (F)~\cite{Ki95, PoWo09, NiCh10, GeTuCi19, LiTo20a, LiTo20b, LuBaCi21, DoLiTo22, ThCl23, DoLi24, IrBaCi24, KaChMe25} offer us alternatives to AQC for ground state preparation.
One such filtering algorithm is the so-called quantum eigenvalue transformation of unitary matrices (QETU)~\cite{DoLiTo22} which employs a polynomial to approximate a step function via quantum singular value transformation (QSVT)~\cite{LoCh17, GiEtAl19, LoCh19, MaEtAl21}.
The QETU method can be used to filter out excited states in the eigenbasis of the target Hamiltonian $H_{\text{targ}}$ for an arbitrary input state $\ket{\psi_{\text{input}}}$, leaving behind an approximation to the target ground state $\ket{\psi_{\text{targ}}}$.
State-of-the-art filter-based algorithms can achieve near-optimal asymptotic complexities $\tilde{O}\left(\Delta_{\text{targ}}^{-1}\right)$ where $\tilde{O}$ denotes the complexity up to polylogarithmic factors and $\Delta_{\text{targ}}$ is the spectral gap of the target Hamiltonian~\cite{GeTuCi19, LiTo20b, KaChMe25}.
Therefore, eigenstate-filtering schemes for ground state preparation can have significantly better asymptotic gate requirements compared to AQC.

Despite their promising asymptotic scaling, filter-based quantum algorithms such as QETU typically require deep quantum circuits to be implemented in practice.
This is due to the sizable constant gate overhead required to realize the controlled time-evolution operator in QETU and the large number of queries to this operator needed to create a sufficiently high-degree polynomial approximation of the desired filter.
These practical challenges are compounded by the probabilistic nature of the QETU scheme in which the filter is realized with a success probability that is low when the input state $\ket{\psi_{\text{input}}}$ has only a small overlap with the target ground state.

In this paper, we aim to address the practical challenges of implementing ground-state filters on digital quantum computers with limited circuit depths.
To that end, we precondition the input state using adiabatic state preparation, $\ket{\psi_{\text{input}}} = \ket{\psi_{\text{AQC}}}$, such that the resource requirements for the filtering algorithm are significantly reduced.
We refer to the combined approach as AQC+F.
Similar strategies have been put forward, from a theoretical perspective, by~\cite{WeEtAl15, WaKi22}.
Here we focus our efforts on experimentally demonstrating the advantage of AQC+F over AQC alone and, in doing so, develop a novel spectral-profiling method which is appropriate for current quantum computers.

\begin{figure}
\centering
\includegraphics[width=0.99\linewidth]{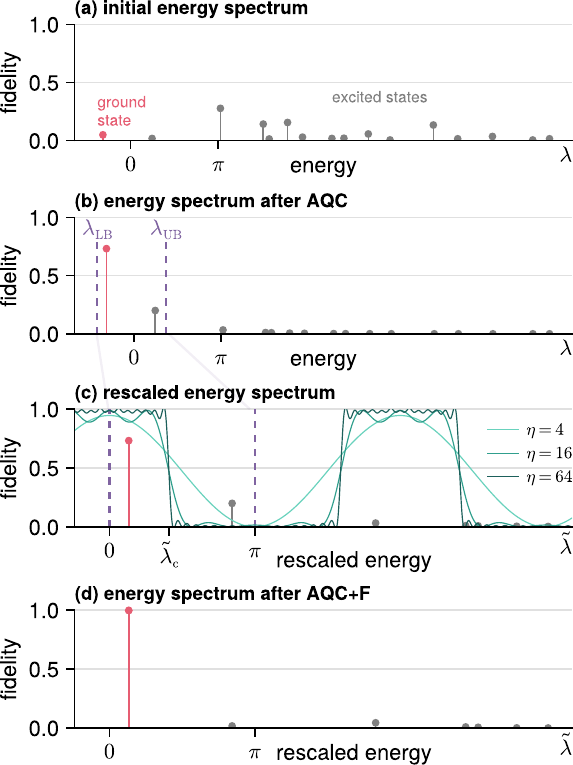}
\caption{\label{fig:summary}
Steps of the AQC+F procedure to prepare the ground state of the target Hamiltonian $H_{\text{targ}}$.
The stages of the procedure are described in terms of the trial state's fidelity spectrum in the eigenbasis of $H_{\text{targ}}$ at each stage, i.e.\ we plot the fidelity $f_{j}(\ket{\varphi}) = \abs{\braket{\varphi}{\psi_{j}}}^2$ for each eigenstate $\ket{\psi_{j}}$ as a function of its energy (rescaled energy) $\lambda_{j}$ ($\tilde{\lambda}_{j}$).
(a) Fidelity spectrum $f_{j}(\ket{\psi_{\text{init}}})$ of the ground state of $H_{\text{init}}$ showing significant overlap with many excited states of $H_{\text{targ}}$.
(b) Fidelity spectrum of the state $\ket{\psi_{\text{AQC}}}$ resulting from adiabatic evolution according to the interpolating Hamiltonian~\eqref{eq:adiabatic_ham}.
Spectral profiling is performed on $\ket{\psi_{\text{AQC}}}$ to determine upper and lower bounds $\lambda_{\text{UB}}$ and $\lambda_{\text{LB}}$, respectively.
(c) Fidelity spectrum $\tilde{f}_{j}(\ket{\psi_{\text{AQC}}})$ with respect to the rescaled Hamiltonian $\tilde{H}$ given in Eq.~\eqref{eq:linear_trafo} where $\lambda_{\text{min}}$ ($\lambda_{\text{max}}$) is $\lambda_{\text{LB}}$ ($\lambda_{\text{UB}}$).
Superimposed are degree $\eta/2$ polynomial functions $\mathcal{F}\left(\tilde{\lambda}\right)$ applied via the QETU circuit according to Eq.~\eqref{eq:qetu_sequence}.
The QETU circuit is chosen such that $\mathcal{F}\left(\tilde{\lambda}\right)$ approximates a periodic square wave filter with discontinuities at $2 l \pi  \pm \tilde{\lambda}_{\text{c}}$ for integer $l$.
(d) Successful application of the lowest degree ($\eta = 4$) filter on the preconditioned input state $\ket{\psi_{\text{input}}} = \ket{\psi_{\text{AQC}}}$ results in a state $\ket{\psi_{\text{AQC+F}}}$ with a large ground-state overlap and strongly suppressed excited states.
}
\end{figure}

We summarize our protocol in Fig.~\ref{fig:summary}.
Starting with an initial state $\ket{\psi} = \ket{\psi_{\text{init}}}$, which is the easy-to-prepare ground state of the Hamiltonian $H_{\text{init}}$, we perform adiabatic evolution according to Eq.~\eqref{eq:adiabatic_ham} for a total evolution time $T$ such that the adiabatically prepared state $\ket{\psi_{\text{AQC}}}$ has significant overlaps with only the ground state and low-lying excited states of $H_{\text{targ}}$.
Next we carry out spectral profiling to compute an upper and a lower energy bound on the significantly occupied states in the eigenbasis of $H_{\text{targ}}$.
Rescaling the target Hamiltonian according to these bounds and using $\ket{\psi_{\text{AQC}}}$ as the input state to the filtering algorithm allows us to utilize the lowest-degree polynomial filter to efficiently suppress the remaining excited states such that only the target ground state is left.
The lowest-degree, non-trivial polynomial in the QETU framework corresponds to $\lfloor \eta/2 \rfloor = 2$~\cite{DoLiTo22}.
In practice, this approach can significantly reduce the gate count of the ground-state preparation circuit compared to, alternatively, simply increasing the QETU polynomial degree~\cite{KaChMe25}.

In our experiments on Quantinuum's H1-1 trapped-ion quantum computer, we observe substantially improved energy accuracies using AQC+F compared to just AQC for a quantum Heisenberg spin chain of eight qubits.
Additionally, we numerically simulate AQC and AQC+F without noise for the transverse-field quantum Ising model on square lattices of sizes up to $5 \times 4$.
In this context, our numerical results for the ground-state overlap show an improvement of at least one order of magnitude, assuming sufficiently large circuit depths and comparing AQC+F with AQC at the same two-qubit gate count.

The remainder of the paper is organized as follows.
Section~\ref{sec:Methods} contains descriptions of the methods that are used in this analysis.
We present the results in Sec.~\ref{sec:Results} and conclude with a discussion in Sec.~\ref{sec:Discussion}.
Further technical details are provided in Appendixes~\ref{app:TrotterCircuits},~\ref{app:QETUCircuits} and~\ref{app:ExperimentDetails}.

\section{Methods}
\label{sec:Methods}

In the following, we first describe the AQC implementation in Sec.~\ref{subsec:AdiabaticTimeEvolution}.
Then we explain the QETU-based eigenstate-filtering procedure in Sec.~\ref{subsec:EigenstateFiltering} as well as the corresponding spectral profiling in Sec.~\ref{subsec:SpectralProfiling}.

\subsection{Adiabatic time evolution}
\label{subsec:AdiabaticTimeEvolution}

We implement digitized AQC by time-dependent Hamiltonian simulation using a product formula approach.
More specifically, we discretize continuous time into $S \cdot T$ time steps of equal size $\tau = 1/S$.
For each time $t_{j} = j \cdot \tau$ we assume that the Hamiltonian is time-independent during the associated time step and has the form $H(t_{j})$ of Eq.~\eqref{eq:adiabatic_ham}.
The entire adiabatic evolution is then given by the product $\prod_{j=1}^{ST} \exp\left[-\text{i}\tau H\left(t_{j}\right)\right]$.
We use a Trotter product formula~\cite{HaSu05} to approximate each of the unitary operators $\exp\left[-\text{i}\tau H\left(t_{j}\right)\right]$ via a quantum circuit.
Further details on the specific Trotter circuits used in this paper are given in Appendix~\ref{app:TrotterCircuits}.

\subsection{Eigenstate filtering}
\label{subsec:EigenstateFiltering}

The aim of the filter is to apply the function $\mathcal{F}\left(\tilde{H}\right)$ to an input state $\ket{\psi_{\text{input}}}$ such that
\begin{equation}\label{eq:filter}
 \mathcal{F}\left(\tilde{H}\right) \ket{\psi_{\text{input}}} = \sum_{j} c_{j} \mathcal{F}\left(\tilde{\lambda}_{j}\right) \ket{\psi_{j}},
\end{equation}
where $c_{j} = \braket{\psi_{j}}{\psi_{\text{input}}}$, $\ket{\psi_{j}}$ is the $j$th eigenstate of $H_{\text{targ}}$, $\mathcal{F}\left(\tilde{\lambda}\right) = F\left[\cos(\tilde{\lambda}/2)\right]$ and $F$ is the QSVT polynomial.
The rescaled Hamiltonian $\tilde{H}$ with eigenvalues $\tilde{\lambda}_{j}$ is given by
\begin{equation}\label{eq:linear_trafo}
 \tilde{H} = \frac{\pi \left(H_{\text{targ}} - \lambda_{\text{min}} \mathds{1}\right)}{\lambda_{\text{max}} - \lambda_{\text{min}}},
\end{equation}
where $\mathds{1}$ is the identity operator and $\lambda_{\text{min}} < \lambda_{\text{max}}$ are real numbers.
In accordance with the original QETU construction~\cite{DoLiTo22}, the real polynomial $F(x)$ for $x \in \mathbb{R}$ is chosen to have even parity, maximum degree $\lfloor \eta/2 \rfloor$, and to satisfy $\abs{F(x)} \leq 1$ for all $x \in [-1, 1]$.
Reference~\cite{DoLiTo22} shows that the filter~\eqref{eq:filter} with these properties is obtained by applying the unitary
\begin{equation}\label{eq:qetu_sequence}
 \mathcal{U}_{\text{QETU}}\left(\vec{\phi}\right) = e^{\text{i} \phi_{\eta} X} cU^{\dag} e^{\text{i} \phi_{\eta-1} X} \cdots cU e^{\text{i} \phi_{0} X}
\end{equation}
to the input state plus one additional ancilla qubit and by post-selecting measurement outcomes related the ancilla qubit.
Here $\exp(\text{i} \phi_{j} X)$ is a Pauli-$X$-matrix rotation of angle $\phi_{j}$ acting on the ancilla qubit, $\vec{\phi} = \left(\phi_{0}, \phi_{1}, \ldots, \phi_{1}, \phi_{0}\right) \in \mathbb{R}^{\eta+1}$ is a symmetric sequence of phase factors and $cU$ denotes the controlled time-evolution operator for $U = \exp(-\text{i} \tilde{H})$ controlled by the ancilla qubit.
As shown in~\cite[Theorem 1]{DoLiTo22}, applying $\mathcal{U}_{\text{QETU}}$ to the quantum state $\ket{\psi_{\text{input}}}$ and the ancilla qubit initialized in $\ket{0}_{\text{anc}}$, and successfully measuring the ancilla qubit in the state $\ket{0}_{\text{anc}}$ creates the desired filter,
\begin{equation}\label{eq:qetu_filter}
 \bra{0}_{\text{anc}} \, \mathcal{U}_{\text{QETU}} \ket{0}_{\text{anc}} \otimes \ket{\psi_{\text{input}}} \propto \mathcal{F}\left(\tilde{H}\right) \ket{\psi_{\text{input}}}.
\end{equation}
It is possible to optimize the QSVT phases $\vec{\phi}$ such that $F(x)$ approximates a symmetric, shifted sign function
\begin{equation}\label{eq:step}
 S(x) = \begin{cases}
    0, & \abs{x} \leq \mu\\
    1, & \abs{x} > \mu
\end{cases},
\end{equation}
for $\abs{x} \in \left[0, 1\right]$ and cutoff value $0 \leq \mu \leq 1$~\cite{DoLiTo22}.
Taking into account the cosine transformation resulting from the combination of forward and backward time evolution, the resulting filter $\mathcal{F}\left(\tilde{H}\right) = F\left[\cos(\tilde{H}/2)\right]$ approximates a square-wave function illustrated in Fig.~\ref{fig:summary}~(c) where $\tilde{\lambda}_{\text{c}} = 2\cos^{-1}(\mu)$.
Appendix~\ref{app:QETUCircuits} provides further information on our QETU circuit constructions.

\subsection{Spectral profiling}
\label{subsec:SpectralProfiling}

In~\cite[Theorem 6]{DoLiTo22} an algorithm is outlined for ground state preparation based on the assumption that the entire spectrum of $\tilde{H}$ is contained in $\left[\nu, \pi-\nu\right]$ for some small $\nu > 0$.
For many interesting Hamiltonians, the required polynomial degree $\eta$ of that approach is too demanding for current hardware.
To alleviate the cost, we exploit that we can prepare the filter input state $\ket{\psi_{\text{AQC}}}$ via sufficiently long adiabatic evolution after which this state has significant overlaps only with the ground state and low-energy eigenstates of the target Hamiltonian.
We use this property to rescale the target Hamiltonian such that only those eigenstates are contained in $\left[\nu, \pi-\nu\right]$.
More precisely, in the rescaling transformation~\eqref{eq:linear_trafo} we set $\lambda_{\text{max}} = \lambda_{\text{UB}}$ to be an upper bound on the highest significantly occupied eigenstate and $\lambda_{\text{min}} = \lambda_{\text{LB}}$ to be a lower bound on the ground-state energy.
The procedure is illustrated in Fig.~\ref{fig:summary}~b--d where one can see that, crucially, even a low-degree polynomial can successfully filter out the remaining excited states.
The purpose of the following method, which we call spectral profiling, is to determine appropriate values for $\lambda_{\text{LB}}$ and $\lambda_{\text{UB}}$.

To perform spectral profiling, we make use of our ability to choose the rescaling parameters $\lambda_{\text{min}}$ and $\lambda_{\text{max}}$ associated with $\tilde{H}$ in Eq.~\eqref{eq:linear_trafo} together with the filter described above, i.e.\ we use the QETU filter $\mathcal{F}\left(\tilde{H}\right)$~\eqref{eq:qetu_filter} where $F(x)$ approximates the shifted sign function of Eq.~\eqref{eq:step}.
Given an input state $\ket{\psi_{\text{input}}}$, the key quantity we are interested in is the success probability --- the probability of measuring $\ket{0}_{\text{anc}}$ after applying the filter --- for a given pair of rescaling parameters $\lambda_{\text{max}}$ and $\lambda_{\text{min}}$.
This success probability depends on the overlaps of $\ket{\psi_{\text{input}}}$ with the eigenstates corresponding to eigenvalues $\tilde{\lambda}$ for which $\mathcal{F}\left(\tilde{\lambda}\right) > 0$, according to~\eqref{eq:qetu_filter}.

\begin{figure}
\centering
\includegraphics[width=0.99\linewidth]{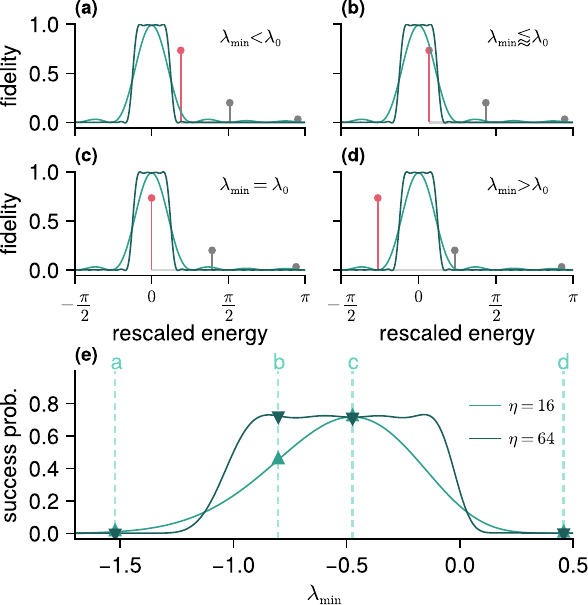}
\caption{\label{fig:lower_bound_scan}
Example of spectral profiling to determine a lower bound $\lambda_{\text{LB}}$ using a filter with $\mu = 0.98$ and $\eta = 16$ or $\eta = 64$, where only the ground and first excited state of $H_{\text{targ}}$ are significantly occupied.
Panels a--d show the spectrum (fidelity as a function of rescaled energy $\tilde{\lambda}$) in the eigenbasis of $\tilde{H}$~\eqref{eq:linear_trafo} for fixed $\lambda_{\text{max}} \gg \lambda_{0}$ and a range of $\lambda_{\text{min}}$, where $\lambda_{0}$ is the true ground state energy of $H_{\text{targ}}$.
In each case, filters $\mathcal{F}$~\eqref{eq:qetu_filter} are superimposed.
(e) The success probability as a function of $\lambda_{\text{min}}$ for two different filters.
The lower bound $\lambda_{\text{LB}} \lessapprox \lambda_{0}$ is chosen to the left of the peak, i.e.\ at or in the vicinity of position b.
}
\end{figure}

To determine $\lambda_{\text{LB}}$ we begin by fixing $\lambda_{\text{max}} \gg \lambda_{0}$, where $\lambda_{0}$ is the true ground state energy of $H_{\text{targ}}$.
We proceed by evaluating the success probability for a range of values of $\lambda_{\text{min}}$ with fixed polynomial degree $\eta$ and cutoff value $\mu$.
Snapshots of the spectrum scan for different values of $\lambda_{\text{min}}$ shown in Figs.~\ref{fig:lower_bound_scan}~a--d demonstrate how the success probability, plotted in Fig.~\ref{fig:lower_bound_scan}~e, is associated to the overlap between the filter and the fidelity spectrum of the input state in the eigenbasis of $\tilde{H}$ and, furthermore, how it varies as a function of $\lambda_{\text{min}}$.

\begin{figure}
\centering
\includegraphics[width=0.99\linewidth]{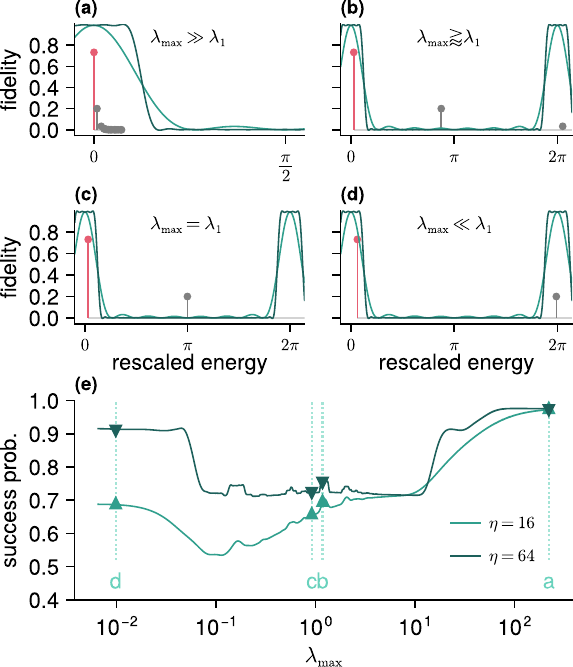}
\caption{\label{fig:upper_bound_scan}
Example of spectral profiling to determine an upper bound $\lambda_{\text{UB}}$ using a filter with $\mu = 0.98$ and $\eta = 16$ or $\eta = 64$, where only the ground and first excited state of $H_{\text{targ}}$ are significantly occupied.
Panels a--d show the spectrum (fidelity as a function of energy) in the eigenbasis of $\tilde{H}$~\eqref{eq:linear_trafo} for different values of $\lambda_{\text{max}}$, where $\lambda_{\text{min}} = \lambda_{\text{LB}}$ is fixed to the value determined from the lower bound scan of Fig.~\ref{fig:lower_bound_scan}, and $\lambda_{1}$ is the true first excited state energy of $H_{\text{targ}}$.
(e) Success probability as a function of $\lambda_{\text{max}}$.
As $\lambda_{\text{max}}$ decreases, dips in the success probability indicate eigenstates being omitted from the filter.
We choose $\lambda_{\text{UB}} \gtrapprox \lambda_{1}$, i.e.\ at or near position b.
For $\lambda_{\text{max}} \ll \lambda_{1}$ the spectral gap acquires a value of $\approx 2 \pi$ at which point the success probability increases due to the periodicity of the filter as shown in panel d.
}
\end{figure}

In the second stage of the spectral profiling we set $\lambda_{\text{min}} = \lambda_{\text{LB}}$ and evaluate the success probability as a function of $\lambda_{\text{max}}$.
Figure~\ref{fig:upper_bound_scan} shows the success probability as a function of $\lambda_{\text{max}}$, as well as a set of snapshots of the spectral profiling procedure used to determine $\lambda_{\text{UB}}$.

It is important to emphasize that the spectral profiling circuits are separate to the final ground state preparation circuits.
In practice, we use larger-degree polynomial filters (and therefore deeper circuits) in spectral profiling than we do in the final ground state preparation.
This is possible because the single-qubit success probability estimation required for spectral profiling is typically more resilient to noise than the multi-qubit energy expectation value estimation that is used as a figure of merit for final ground state preparation.

\section{Results}
\label{sec:Results}

We now analyze how AQC+F performs, compared to pure AQC, in the context of two paradigmatic quantum spin models.
We consider ground state preparation, firstly, for a one-dimensional (1D) quantum Heisenberg Hamiltonian in Sec.~\ref{subsec:1DHeisenbergModel} and, secondly, for two-dimensional (2D) transverse-field Ising Hamiltonians in Sec.~\ref{subsec:2DTransverse-FieldIsingModel}.

\subsection{1D Heisenberg model}
\label{subsec:1DHeisenbergModel}

We demonstrate the AQC+F protocol by preparing the ground state of a one-dimensional Heisenberg model (HM) with periodic boundary conditions.
The target Hamiltonian is given by
\begin{equation}\label{eq:Heisenberg_targ}
 H^{\text{(HM)}}_{\text{targ}} = \sum_{\mathcal{O} \in \{X, Y, Z\}} \sum_{j=0}^{L-1} \mathcal{O}_{j} \mathcal{O}_{j+1}
\end{equation}
where $X$, $Y$ and $Z$ are the Pauli matrices, $\mathcal{O}_{L} = \mathcal{O}_{0}$ due to the periodic boundary conditions, and we choose the system size $L$ to be an even number.
We initialize the system in the ground state of the initial Hamiltonian
\begin{equation}\label{eq:Heisenberg_init}
 H^{\text{(HM)}}_{\text{init}} = \sum_{\mathcal{O} \in \{X, Y, Z\}} \sum_{j=0}^{L/2-1} \mathcal{O}_{2j} \mathcal{O}_{2j+1}.
\end{equation}
The adiabatic path of the interpolating Hamiltonian with total time $T$ is given by
\begin{equation}\label{eq:adiabatic_path}
 s(t) = \sin^{2}\left(\frac{\pi t}{2 T}\right).
\end{equation}
Details of the Trotter circuits used to implement the time-dependent evolution along this path are provided in Appendix~\ref{app:TrotterCircuits}.
We implement the controlled time-evolution operators required for the QETU circuit~\eqref{eq:qetu_sequence} using the so-called control-free implementation~\cite{DoLiTo22}.
This exists for Hamiltonians that can be expressed as a sum of terms such that each term anti-commutes with a specific single Pauli string.
For additional details on how to implement the QETU circuit we refer the reader to Appendix~\ref{app:QETUCircuits}.

\begin{figure}
\centering
\includegraphics[width=0.99\linewidth]{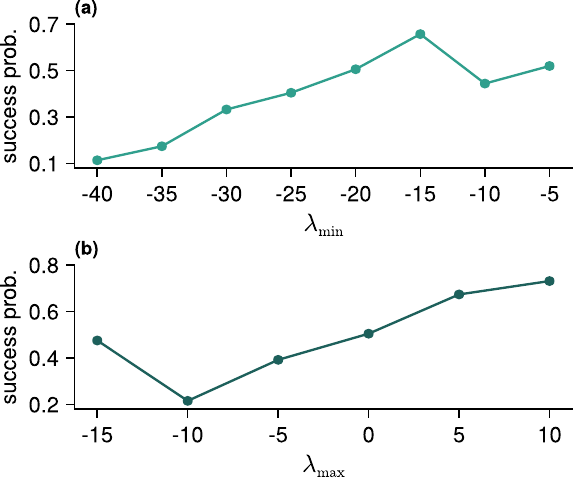}
\caption{\label{fig:spectrum_scan}
Spectral profiling for the Heisenberg model of Eq.~\eqref{eq:Heisenberg_targ} with 8 spins and for total adiabatic time $T = 5$.
We carry out the success probability measurements using $\eta = 12$ and $\mu = 0.8$ on a grid of energy values where $\lambda_{\text{min}} \in [-40, -5]$ (a) and $\lambda_{\text{max}} \in [-15, 10]$ (b).
The results are computed using the emulator of Quantinuum's H1-1 trapped-ion quantum device.
Each data point is calculated from 3000 measurements of the ancilla qubit.
For the lower bound scan (a), we set $\lambda_{\text{max}}$ to $0$.
}
\end{figure}

Figure~\ref{fig:spectrum_scan} shows the results of spectral profiling for the HM with 8 spins after an adiabatic evolution of time $T = 5$, conducted using the emulator of Quantinuum's H1-1 quantum computer taking into account realistic hardware noise.
In Fig.~\ref{fig:spectrum_scan}~(a) we plot the success probability as a function of $\lambda_{\text{min}}$ where the peak centered at $\lambda_{\text{min}} \approx -15$ enables us to determine the lower bound $\lambda_{\text{LB}} = -20$.
Similarly, in Fig.~\ref{fig:spectrum_scan}~(b) we observe dips in the success probability as a function of $\lambda_{\text{max}}$ which allow us to locate the upper bound $\lambda_{\text{UB}} = -5$.
By choosing the value of $\lambda_{\text{LB}}$  ($\lambda_{\text{UB}}$) immediately to the left (right) of the peak (dip) of the success probability, we avoid incorrectly determining a lower (upper) bound due to the finite resolution in $\lambda_{\text{min}}$ ($\lambda_{\text{max}}$).

Using these rescaling parameters, we compare the performance of AQC+F to the one of AQC without any filter by evaluating the respective circuits via both noiseless statevector simulations and via hardware experiments on the Quantinuum H1-1 quantum computer.
In particular, we calculate the relative energy error
\begin{equation}\label{eq:relative_energy_error}
 \epsilon_{\text{E}}(\lambda) = \frac{\abs{\lambda - \lambda_{0}}}{\abs{\lambda_{0}}}
\end{equation}
between its exact value $\lambda_{0}$, calculated numerically, and its measured value $\lambda$.
For further details on the post-processing procedures carried out using the measurement results related to $\lambda$, we refer the reader to Appendix~\ref{app:ExperimentDetails}.

\begin{figure}
\centering
\includegraphics[width=0.99\linewidth]{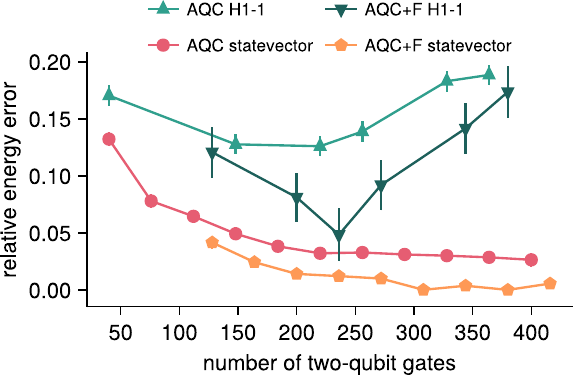}
\caption{\label{fig:main_results}
Relative energy error~\eqref{eq:relative_energy_error} for ground state approximations of the Heisenberg model~\eqref{eq:Heisenberg_targ} with 8 spins as a function of the number of native Quantinuum two-qubit gates $\exp(-\text{i} \phi Z \otimes Z / 2)$ with variable angle $\phi$ contained in the state preparation circuit.
Quantum states are prepared using either AQC or AQC+F.
Results from circuits evaluated via noiseless statevector simulations are compared to those of circuits run on Quantinuum's H1-1 trapped-ion quantum computer.
For the AQC+F results we use the lowest-degree polynomial filter with $\eta = 4$ and the cutoff value $\mu = 0.8$.
For both AQC and AQC+F we use $S = 3$ Trotter steps per unit time.
We vary the adiabatic time $T \in \{1, 4, 6, 7, 9, 10\}$ for AQC and $T \in \{1, 3, 4, 5, 7, 9\}$ for AQC+F.
For the hardware results, energies are estimated using up to 1000 measurement shots per observable and the associated error bars represent an upper bound for the standard deviation of the sampling.
Larger error bars for AQC+F are due to the discarding of shots for which the ancilla qubit is measured in the unwanted state $\ket{1}_{\text{anc}}$.
For the statevector simulations, the energy was computed using 10000 shots per observable.
}
\end{figure}

\begin{figure*}
\centering
\includegraphics[width=0.99\linewidth]{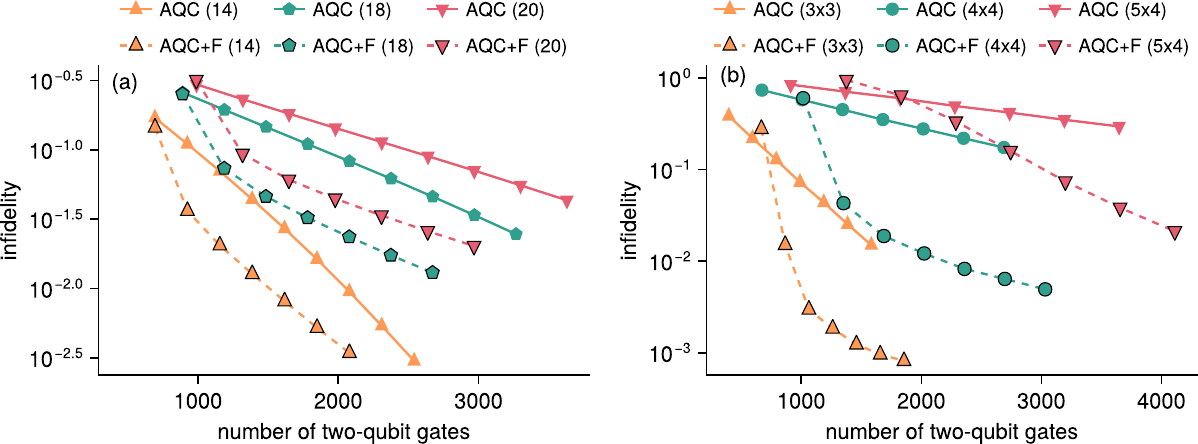}
\caption{\label{fig:2D_results}
Ground-state infidelity $\epsilon_{\text{F}}$~\eqref{eq:infidelity} obtained by AQC and AQC+F for Heisenberg chains~\eqref{eq:Heisenberg_targ} of lengths $L \in \{14, 18, 20\}$ (a) and for transverse-field Ising models~\eqref{eq:Ising_targ} of various sizes $\left(L_{x} \times L_{y}\right)$ (b).
The error is plotted as a function of the number of generic two-qubit gates.
Results are calculated using statevector simulations.
For the AQC+F results, we use the lowest-order $\eta = 4$ QETU filter with cutoff value $\mu = 0.8$, which, for each system size, contributes a constant number of two-qubit gates.
We realize the digitized adiabatic time evolution circuits using the fourth-order product formula~\eqref{eq:TrotterIV} for the Heisenberg model and the second-order product formula-\eqref{eq:Trotter} for the Ising model.
For both AQC and AQC+F, we use $S = 3$ Trotter steps per unit time and evaluate circuits by means of statevector methods.
We extract the spectrum bounds for AQC+F via spectral profiling using statevector simulations where $\mu = 0.98$, $\eta = 12$ and energy grids have unit spacing ranging from $-100$ to $0$ (for both $\lambda_{\text{LB}}$ and $\lambda_{\text{UB}}$ scans).
}
\end{figure*}

Figure~\ref{fig:main_results} shows $\epsilon_{\text{E}}(\lambda)$ for the HM~\eqref{eq:Heisenberg_targ} with 8 spins as a function of the total number of two-qubit gates used in each state preparation circuit.
For the AQC as well as the AQC+F results, those circuits with larger two-qubit gate counts realize a larger total adiabatic time $T$, where, for the AQC+F results, the number of two-qubit gates contributed by the filter remains the same throughout (corresponding to 88 native Quantinuum two-qubit gates $\exp(-\text{i} \phi Z \otimes Z / 2)$ with variable angle $\phi$).
In both the statevector and hardware results of Fig.~\ref{fig:main_results}, we observe that AQC+F consistently outperforms AQC for all total adiabatic evolution times investigated.
We also see in Fig.~\ref{fig:main_results} that increasing the adiabatic evolution time, and thereby the number of two-qubit gates, reduces the energy error for statevector results as expected.
For the hardware results, however, energy errors begin to increase for large adiabatic evolution times due to the accumulation of hardware noise resulting from the larger circuit depths.

In addition to the experimental analysis of the 8-spin HM, let us also consider larger system sizes via statevector simulations.
To compare the performance of AQC and AQC+F, we compute the state infidelity
\begin{equation}\label{eq:infidelity}
 \epsilon_{\text{F}}(\ket{\varphi}) = 1 - \abs{\braket{\varphi}{\psi_{\text{targ}}}}^{2}
\end{equation}
between the true ground state $\ket{\psi_{\text{targ}}}$, calculated numerically, and the state $\ket{\varphi}$ prepared either by AQC or AQC+F.
Our results are shown in Fig.~\ref{fig:2D_results} a.
We see that, for sufficiently deep circuits, AQC+F consistently outperforms AQC.

\subsection{2D transverse-field Ising model}
\label{subsec:2DTransverse-FieldIsingModel}

Next we apply the AQC+F protocol to the preparation of ground states of a transverse-field Ising model (TFIM) on two-dimensional square lattices of size $L_{x} \times L_{y}$ with periodic boundary conditions.
The target Hamiltonian is given by
\begin{equation}\label{eq:Ising_targ}
 H^{(\text{TFIM})}_{\text{targ}} = \sum_{j=0}^{L_{x}L_{y}-1} X_{j} - \sum_{\langle j, k \rangle} Z_{j} Z_{k}
\end{equation}
where the sum over $\langle j, k \rangle$ runs over all pairs of sites $j$ and $k$ that are nearest neighbors on the square lattice.
We initialize the system in the ground state of the Hamiltonian
\begin{equation}\label{eq:Isisng_init}
 H^{(\text{TFIM})}_{\text{init}} = \sum_{j=0}^{L_{x}L_{y}-1} X_{j}
\end{equation}
and for the adiabatic path we use~\eqref{eq:adiabatic_path}.

In Fig.~\ref{fig:2D_results} b the infidelity $\epsilon_{\text{F}}$ is plotted as a function of the number of two-qubit gates contained in the state preparation circuits.
We observe that, except for the lowest two-qubit gate counts considered, AQC+F produces significantly reduced infidelities compared to AQC at comparable gate counts, with an improvement of at least one order of magnitude.
We conjecture that, for the problems studied in Fig.~\ref{fig:2D_results}, the input state to the filter, $\ket{\psi_{\text{AQC}}}$, significantly suppresses excited states already after relatively short adiabatic evolution times, i.e.\ corresponding to a low number of two-qubit gates in Fig.~\ref{fig:2D_results}.
This would explain why the advantage of AQC+F over AQC is visible already for low two-qubit gate counts in Fig.~\ref{fig:2D_results}.

\section{Discussion}
\label{sec:Discussion}

For paradigmatic one- and two-dimensional quantum spin models, we have experimentally and numerically demonstrated that AQC+F, i.e.\ the combination of adiabatic quantum computing (AQC) with eigenstate filtering (F), can prepare better ground states than AQC.
Excluding experimental noise, we have numerically shown that AQC+F leads to systematically improved ground state results when the circuit depth increases.
In the presence of experimental noise, we observe that, when the circuit depth increases, the ground state accuracy first improves and then worsens (due to the accumulation of noise in deeper circuits), such that there exists an optimal circuit depth.
For the realization of AQC+F on current quantum computers without error correction, it is therefore interesting to develop a protocol to determine the optimal circuit depth in analogy to~\cite{KiEtAl23}.

The performance of AQC+F is sensitive to both the minimum spectral gap along the adiabatic path and the spectral gap of the target Hamiltonian.
Smaller spectral gaps lead to deeper circuits for multiple reasons:
They require longer adiabatic evolution times, larger-degree polynomials for spectral profiling and longer effective evolution times in the controlled unitary evolutions of the QETU circuits.
Regarding the performance of AQC+F in the presence of gate noise, we expect that, as spectral gaps of target problems decrease, the maximum achievable advantage of AQC+F over AQC decreases.

It is worth emphasizing that AQC+F can readily be combined with circuit compression techniques and will likely benefit from them.
For adiabatic evolution and Hamiltonian simulation, several quantum circuit compression algorithms have been developed and shown to produce circuits that can significantly outperform their counterparts derived from Trotter product formulas~\cite{MaEtAl23, TeHaLu23, McLu23, MaFiHa23, KoEtAl24, McLu24}.
These algorithms can directly be utilized here to compress the AQC circuit block as well as the time-evolution operator of QETU.

To improve the performance of the AQC circuit, a powerful toolset to consider is provided by so-called shortcuts to adiabaticity~\cite{HeEtAl21a}.
The latter concept has proven to be successful in numerous applications~\cite{HeEtAl21b, ChEtAl22, RoEtAl24, CaEtAl25}.

Throughout our experiments, we rely on expectation value measurements via sampling to approximate the ground state energy.
To get better energy accuracies, we can use a dedicated circuit layer for the energy extraction at the end of ground state preparation, e.g.\ through iterative quantum phase estimation (IQPE) algorithms or similar approaches~\cite{LiTo22, To22, DiLi23a, DiLi23b, NiLiYi23, LiNiYi23, DiEtAl24}.
The IQPE methods estimate the energy one digit after another.
They come with the additional cost of one more query to the time-evolution operator, whose simulation time depends on the targeted digit.
Mitigating the effects of hardware noise on these schemes is another interesting future research direction.

\appendix

\section{Trotter circuits}
\label{app:TrotterCircuits}

\begin{figure}
\centering
\includegraphics[width=0.95\linewidth]{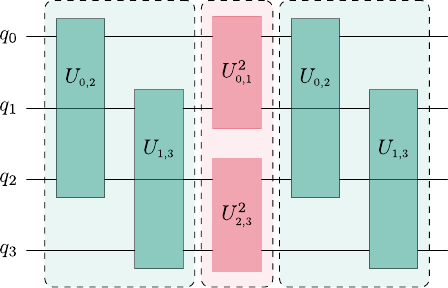}
\caption{\label{fig:tfim_2D_trotter}
Illustration of the Trotter circuit used in this paper to implement a single time step of AQC for a $2 \times 2$ lattice.
We implement the time-evolution operator $\exp(-\text{i} \tau H_{j})$ of each instantaneous Hamiltonian $H_{j}$ using the Trotter formula~\eqref{eq:Trotter}.
The circuit is composed of gates $U_{k, l} = \exp(-\text{i} \tau H^{\text{loc}, j}_{k, l})$ where $H^{\text{loc}, j}_{k, l}$ is defined in Eq.~\eqref{eq:HlocTFIM}.
}    
\end{figure}

We approximate the time evolution according to the Schr\"{o}dinger equation with the time-dependent Hamiltonian $H(t)$~\eqref{eq:adiabatic_ham}, where the adiabatic path is defined in~\eqref{eq:adiabatic_path}, for a total time $T$ by a product of time-independent evolutions governed by the Hamiltonians $H_{j} = H(j \tau)$ where $\tau = 1/S$ and $j = 1, 2, \dots ST$.
For both the transverse field Ising model (TFIM) and the Heisenberg model (HM), each time-independent step is approximated using either the second-order Trotter-Suzuki product formula~\cite{Su91}
\begin{equation}\label{eq:Trotter}
 \mathscr{S}_{2}(\tau) = \left( \prod_{\langle k, l \rangle} e^{-\frac{\text{i}\tau}{2} H^{\text{loc}, j}_{k, l}} \right) \left( \prod_{\langle k, l \rangle} e^{\frac{\text{i}\tau}{2} H^{\text{loc}, j}_{k, l}} \right)^{\dag},
\end{equation}
or the fourth-order product formula~\cite{Su91}
\begin{equation}\label{eq:TrotterIV}
 \mathscr{S}_{4}(\tau) = \left[\mathscr{S}_{2}(u \tau)\right]^{2} \mathscr{S}_{2}\left((1-4u)\tau\right) \left[\mathscr{S}_{2}(u \tau)\right]^{2},
\end{equation}
where $u = 1 / \left(4 - 4^{1/3}\right)$.
Here $\langle k, l \rangle$ denotes nearest neighbors on the lattice, $H^{\text{loc}, j}_{k, l}$ is the term from $H_{j}$ acting on qubits $k$ and $l$, and the product is ordered in groups such that all gates per group commute by acting on different qubit pairs.
The second-order scheme $\mathscr{S}_{2}(\tau)$ of Eq.~\eqref{eq:Trotter} is illustrated in Fig.~\ref{fig:tfim_2D_trotter} for the case of a $2 \times 2$ lattice.

For the two-dimensional TFIM we define
\begin{equation}\label{eq:HlocTFIM}
 H^{\text{loc}, j}_{k, l} = -s(j \tau) Z_{k} Z_{l} + \frac{1}{4} \left( X_{k} + X_{l} \right)
\end{equation}
where $s(t)$ goes from $0$ to $1$ according to Eq.~\eqref{eq:adiabatic_path}.
The resulting evolution slowly turns on the interaction strength such that the system interpolates between the initial Hamiltonian $H^{\text{TFIM}}_{\text{init}}$~\eqref{eq:Isisng_init} and the final Hamiltonian $H^{\text{TFIM}}_{\text{targ}}$~\eqref{eq:Ising_targ}.

\begin{figure*}
\centering
\includegraphics[width=0.99\linewidth]{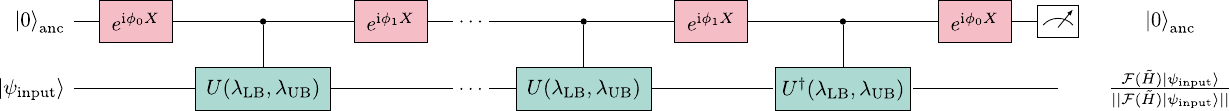}
\caption{\label{fig:qetu_circuit}
Circuit representation corresponding to quantum eigenvalue transformation of unitary matrices (QETU)~\cite{DoLiTo22}.
Here $U = \exp(-\text{i}\tilde{H})$ is the unitary operator realizing time evolution with the rescaled Hamiltonian $\tilde{H}$ given in Eq.~\eqref{eq:linear_trafo} where ($\lambda_{\text{min}}, \lambda_{\text{max}}$) are fixed to ($\lambda_{\text{LB}}, \lambda_{\text{UB}}$).
The operator $U$ acts on the system qubits.
The single-qubit rotation gates $\exp(\text{i} \phi_{j} X)$ with rotation angles $\phi_{j}$ are applied to the ancilla qubit.
The symmetric phases $\left(\phi_{0}, \phi_{1}, \ldots, \phi_{1}, \phi_{0}\right) \in \mathbb{R}^{\eta+1}$ are optimized according to the desired target polynomial $\mathcal{F}\left(\tilde{H}\right)$.
}
\end{figure*}

For the one-dimensional HM we define
\begin{equation}\label{eq:HlocHM}
 H^{\text{loc}, j}_{k, l} = \sum_{\mathcal{O} \in \{X, Y, Z\}} \mathcal{O}_{k} \mathcal{O}_{l}\left([s(j \tau)-1] \delta_{k\%2, 1} + 1\right)
\end{equation}
where $\delta_{k\%2,1}$ represents a Kronecker delta which assumes the value $0$ when $k$ is divisible by $2$.
The resulting evolution starts from the product of singlet states and then slowly turns on interactions between the sites $\{(2j+1, 2j+2)\}_{j=0}^3$, which slowly interpolates between the initial Hamiltonian $H^{\text{HM}}_{\text{init}}$~\eqref{eq:Heisenberg_init} and the target Hamiltonian $H^{\text{HM}}_{\text{targ}}$~\eqref{eq:Heisenberg_targ}.

\section{QETU circuits}
\label{app:QETUCircuits}

The circuit implementing the QETU filter is depicted in Fig.~\ref{fig:qetu_circuit} and consists of the controlled unitary operator $U = \exp(-\text{i}\tilde{H})$ and its conjugate $U^{\dag} = \exp(\text{i}\tilde{H})$ as well as single-qubit rotation gates acting on the ancilla qubit initialized in the state $\ket{0}_{\text{anc}}$.
The rescaled target Hamiltonian $\tilde{H}$ corresponds to Eq.~\eqref{eq:linear_trafo} with $(\lambda_{\text{min}}, \lambda_{\text{max}})$ set to $(\lambda_{\text{LB}}, \lambda_{\text{UB}})$ which are determined during the spectral profiling stage explained in Sec.~\ref{subsec:SpectralProfiling}.
We realize the controlled unitary operators via the so-called control-free approach outlined in~\cite{DoLiTo22} and use the Trotterization given in Eq.~\ref{eq:Trotter} to implement the time evolution operator $U$, with $S = \lceil \frac{\pi}{0.1(\lambda_{\text{UB}} - \lambda_{\text{LB}})} \rceil$ and $H_{j}$ set to $\tilde{H}$, whose local Hamiltonians $\tilde{H}^{\text{loc}}_{k, l}$ are the same as given in Eq.~\ref{eq:HlocTFIM} (Eq.~\ref{eq:HlocHM}) for TFIM (for HM) with $s(j \tau)$ set to $1$. More explicitly, we use
\begin{equation}\label{eq:TrotterQETU}
 e^{-\text{i}\tilde{H}} \approx \left[ \left( \prod_{\langle k, l \rangle} e^{-\frac{\text{i}}{2S} \tilde{H}^{\text{loc}}_{k, l}} \right) \left( \prod_{\langle k, l \rangle} e^{\frac{\text{i}}{2S} \tilde{H}^{\text{loc}}_{k, l}} \right)^{\dag} \right]^{S}.
\end{equation}

When multiple Trotter layers are required in the time evolution operator, the control-free implementation allows for the combination or cancellation of the associated controlled Pauli strings.
While rescaling the Hamiltonian may necessitate multiple Trotter layers, this combination or cancellation of Pauli strings means that gate counts for the rescaled approach, with fixed polynomial degree, are typically more favorable than the alternative approach which involves increasing the polynomial degree.

We determine the symmetric phase factors $\vec{\phi} = \left(\phi_{0}, \phi_{1}, \ldots, \phi_{1}, \phi_{0}\right)$ with the help of the python library \texttt{pyqsp}~\cite{ChEtAl20, DoEtAl21, MaEtAl21}.

\section{Experiment details}
\label{app:ExperimentDetails}

We conducted the hardware experiments using the Quantinuum H1-1 trapped-ion quantum computer on the 23rd of August 2024.
We compiled our circuits using the \texttt{pytket} compiler with the default settings.
We estimate the energy of the prepared state $\ket{\varphi}$ through expectation value measurements via sampling.
For the HM we calculate the energy using the estimated probability distributions as
\begin{equation}
 \lambda = \sum_{\mathcal{O} \in \{X, Y, Z\}} \sum_{\langle k, l \rangle} \bra{\varphi} \mathcal{O}_{k} \mathcal{O}_{l} \ket{\varphi}
\end{equation}
where $\langle k, l \rangle$ sums over the nearest neighbors and the expectation value $\bra{\varphi} \mathcal{O}_{k} \mathcal{O}_{l} \ket{\varphi}$ is defined as
\begin{equation}
 \bra{\varphi} \mathcal{O}_{k} \mathcal{O}_{l} \ket{\varphi} = \sum_{\sigma_{k}, \sigma_{l}} (-1)^{\sigma_{k}}  (-1)^{\sigma_{l}} P_{\mathcal{O}} \left( \sigma_{k}, \sigma_{l}| \varphi \right).
\end{equation}
Here $P_{\mathcal{O}} \left( \sigma_{k}, \sigma_{l} | \varphi \right)$ is the probability of measuring the $k^{th}$ and $l^{th}$ qubits in the state $\ket{\sigma_{k}} \ket{\sigma_{l}}$ in the $\mathcal{O}$-basis and $(\sigma_k, \sigma_l) \in \{(0,0), (0,1), (1,0), (1,1)\}$.
These probabilities were estimated by 1000 shots on the quantum hardware for each $\mathcal{O}$-basis.

Error bars of our results are taken as an upper bound for the standard deviation of the sampling results.
In particular, we note that the variance, i.e.\ the squared standard deviation, satisfies
\begin{align}
 \text{Var}\left(\mathcal{O}_{1}+\mathcal{O}_{2}\right) &= \text{Var}\left(\mathcal{O}_{1}\right) + \text{Var}\left(\mathcal{O}_{2}\right) + 2 \text{Cov}\left(\mathcal{O}_{1}, \mathcal{O}_{2}\right)\\ &\leq \label{eq:statistical_upper_bound}
 \text{Var}\left(\mathcal{O}_{1}\right) + \text{Var}\left(\mathcal{O}_{2}\right)
\end{align}
since $\text{Cov}\left(\mathcal{O}_{1}, \mathcal{O}_{2}\right) \leq 0$ for anti-commuting Pauli strings $\mathcal{O}_{1}$ and $\mathcal{O}_{2}$~\cite{DaDeJa00}.
Therefore, as an upper bound for the standard deviation we define $\delta \lambda_{\text{max}}$ via
\begin{equation}
 \delta \lambda^{2} \leq \delta \lambda^{2}_{\text{max}} = \sum_{\mathcal{O} \in \{X, Y, Z\}} \sum_{\langle k, l \rangle} \frac{1 - \bra{\varphi} \mathcal{O}_{k} \mathcal{O}_{l} \ket{\varphi}^{2}}{M}.
\end{equation}
Here we use $\bra{\varphi} \left( \mathcal{O}_{k} \mathcal{O}_{l} \right)^{2} \ket{\varphi} = 1$ for any Pauli string $\mathcal{O}_{k} \mathcal{O}_{l}$ and $M$ is the number of samples.
In our experiments $M$ is equal to 1000 for AQC but reduced for AQC+F due to the post-selection of measuring the ancilla qubit in $\ket{0_{\text{anc}}}$.
The reduction in $M$ leads to larger error bars for the AQC+F results compared with the AQC results, visible in Fig. \ref{fig:main_results}.

\bibliography{bibliography.bib}

\end{document}